# Mass extinctions and supernova explosions

**Gunther Korschinek**

Physik-Department, Technische Universität München

korschin@tum.de

**Abstract**

A nearby supernova (SN) explosion could have negatively influenced life on Earth, maybe even been responsible for mass extinctions. Mass extinction poses a significant extinction of numerous species on Earth, as recorded in the paleontologic, paleoclimatic, and geological record of our planet. Depending on the distance between the Sun and the SN, different types of threats have to be considered, such as ozone depletion on Earth, causing increased exposure to the Sun's ultraviolet radiation, or the direct exposure of lethal x-rays. Another indirect effect is cloud formation, induced by cosmic rays in the atmosphere which result in a drop in the Earth's temperature, causing major glaciations of the Earth. The discovery of highly intensive gamma ray bursts (GRBs), which could be connected to SNe, initiated further discussions on possible life-threatening events in Earth's history. The probability that GRBs hit the Earth is very low. Nevertheless, a past interaction of Earth with GRBs and/or SNe cannot be excluded and might even have been responsible for past extinction events.

## 1 Introduction

In 1912, V. Hess "(Hess 1912)" discovered cosmic rays. Utilizing a so called Wulfscher Strahlungsapparat "(Wulfscher Strahlungsapparat 2015)" during several balloon rides for the measurements, he found that background radioactivity increases with rising altitude. In 1936 he received the Nobel prize for his discovery. It took 22 years until work by W.Baade and F. Zwicky "(W.Baade and F. Zwicky, 1934)" indicated that fluctuations in cosmic ray intensity can be caused by supernovae (SNe), a word also coined by them. SNe are among the most energetic events in the universe, occurring either after the core-collapse of massive stars running out of nuclear fuel (type II SNe), or in binary systems, where mass flow from a main sequence star onto a white dwarf, exceeding the Chandrasekhar limit and causing a thermonuclear explosion (type Ia SNe).

Cosmic rays are a particle radiation, composed mostly of protons and a small fraction of heavier atomic nuclei and electrons. The sources of these particles include our sun and SN explosions. In addition there are photons, penetrating into the solar system, including x-rays and gamma rays. An interesting question arose, whether a nearby SN explosion could have negative influence on life on Earth, maybe even be responsible for mass extinctions. The term mass extinction refers to a significant extinction of numerous species on Earth, as recorded in the paleontologic, paleoclimatic, and geological record of our planet.

Twenty years after the discovery of cosmic rays, one of the earliest speculations about the connection between mass extinctions on Earth and SN explosions was presented by O.H. Schindewolf "(Schindewolf 1954)". He assumed that highly energetic radiation from a SN might cause the extinction of marine organisms, either by the radiation directly or indirectly by forming hazardous radioisotopes. Since then, different authors "(Terry and Tucker 1968)", "(Rudermann 1975)", "(Ellis and Schramm 1995)", and later Gehrels et al. "(Gehrels et al. 2003) studied further aspects of correlations between SNe and mass extinctions.

Figure 1 depicts the fraction of genera that are present in each interval of time but do not exist in the following interval. The yellow line is a cubic polynomial to show the long-term trend. Note that these data do not represent all genera that have ever lived, but rather only a selection of marine genera whose qualities are such that they are easily preserved as fossils.

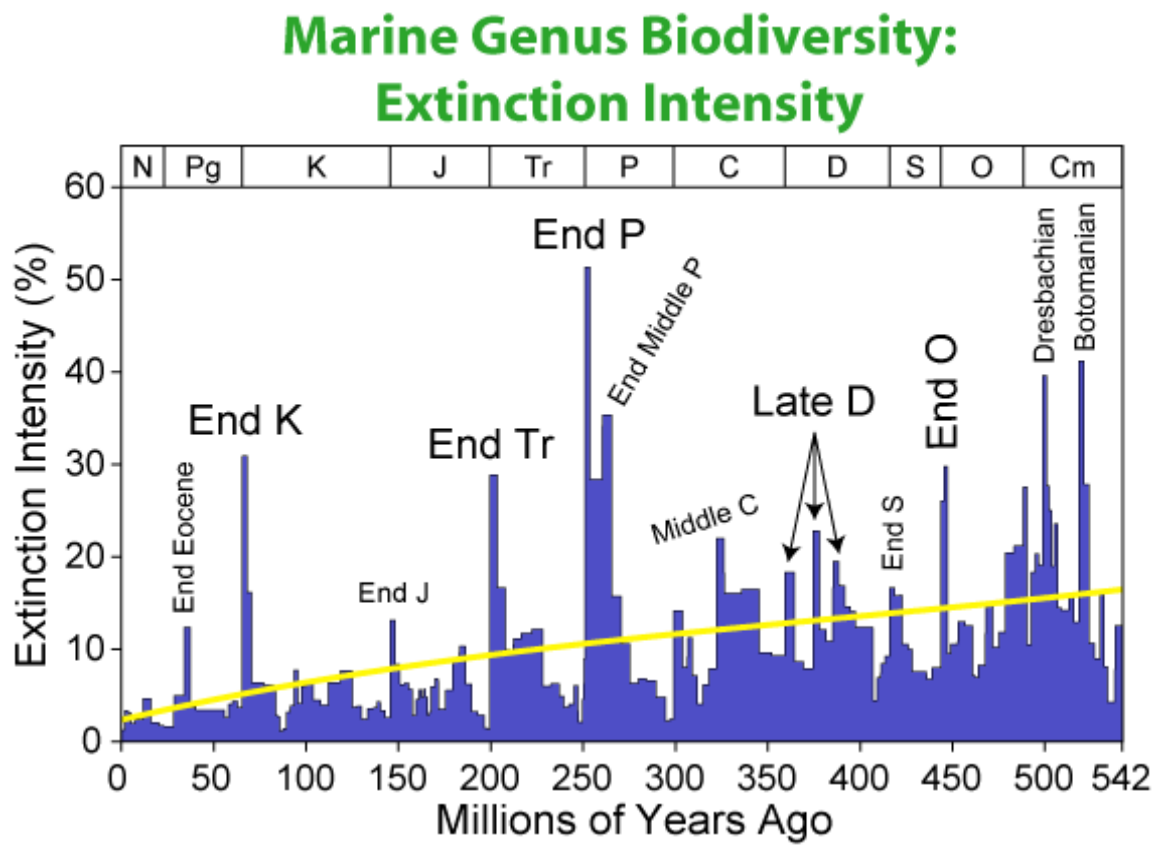

FIG.1

This figure shows the fraction of genera that are present in each interval of time but do not exist in the following interval. The data itself is taken from Rohde and Muller (2005, Supplementary Material) "(Rohde and Muller (2005)" (Different cycle lengths have been proposed; e.g. by Rohde, R.A. and Muller, R. A., and are based on the Sepkoski's Compendium of Marine Fossil Animal Genera "(Sepkoski 2002)". The yellow line is indicating the long-term trend. Note that these data do not represent all genera that have ever lived, but rather only a selection of marine genera whose qualities are such that they are easily preserved as fossils. The figure, part of the accompanying wording, and the caption (shortened) are taken from GNU Free Documentation License. This version is released under the GFDL.

The "Big Five" mass extinctions "(Raup and Sepkoski 1982)" are labeled in large font, and a variety of other features are labeled in smaller font. The two extinction events occurring in the Cambrian (i.e. Dresbachian and Botomian) are very large in percentage magnitude, but are not well known because of the relative scarcity of fossil producing life at that time (i.e. they are small in absolute numbers of known taxa). The Middle Permian extinction is now argued by many to constitute a distinct extinction horizon, though the actual extinction amounts are sometimes lumped together with the End Permian extinctions in literature. As indicated, the "Late Devonian" extinction is actually resolvable into at least three distinct events spread out over a period of ~40 million years. As these data are derived at the genus level, one can anticipate that the number of actual species extinctions is even larger than shown here.

Such mass extinction events can have palaeoclimatic, palaeoecologic, and palaeoenvironmental explanations "(Twitchett 2006)". For example the KT (Cretaceous-Tertiary) extinction (66Ma B.P., indicated in the graph by **End K**) yielding the famous dinosaur-killing mass extinction "(Alvares et al. 1980)", is thought to have been caused by a massive meteorite impact. However there are still inconsistences which might contradict a large meteorite impact only. A different scenario is discussed in a recent publication by R.R. Large et al. "(Large et al. 2015)"; it is argued that a lack of trace nutrient elements in the Phanerozoic ocean might have caused mass extinctions. An additional scenario for mass extinctions that can be considered, are nearby SN explosions.

**2 Different effects of nearby SNe**

We can roughly distinguish two different scenarios for the effect of nearby SN on the Earth. On the one hand, (case A) Earth can be exposed to an irradiation by cosmic rays, which have travelled through the interstellar medium (ISM). This effect is expected to dominate for SN distances of above 15pc or 20pc, depending on the density of the ISM. The other scenario (B) involves a direct interaction of the SN ejecta in the form of plasma overcoming the solar wind pressure and penetrating deep into the solar system, reaching the Earth.

## 2.1. Case A

In case of the first situation (A), the cosmic ray intensity at different distances and times has been analyzed by Knie et al. "(Knie et al. 2004)".

It has been supposed that a first order Fermi mechanism operating in shock waves is the most promising mechanism for the source of galactic CR. For the numerical simulations "(Dorf 1990)" of a SNR evolution in spherical symmetry, a typical value of $10^{51}$ erg has been assumed for the total ejection energy (not counting neutrino losses). The time-dependent acceleration of CR is included through a hydrodynamical formalism characterizing the cosmic rays where a mean diffusion coefficient $K_{CR}$ = 1027 cm$^2$s$^{-1}$, as well as the adiabatic coefficient $\gamma_{CR}$ = 4/3, had to be specified in accordance with the observed properties of CR. The intensity and the duration of the CR exposure depend on the gas density of the surrounding ISM where Dorfi "(Dorfi 1990)" has chosen values between 0.1 atoms cm$^{-3}$ and 1 atom cm$^{-3}$. During the SNR evolution the total energy of the ejecta is shared between kinetic energy of the ejecta (including CR), thermal energy, and photons. In particular, the SNR evolution at later stages is characterized by radiative losses of the thermal plasma. The onset of this radiative phase depends on the particle density and occurs for the adopted values of the ISM for radii larger than about 20 pc "(Kahn 1976)". Hence, the amount of cosmic rays accelerated by the remnants shock wave can only be calculated if the radiative cooling effects are taken into account. A typical result for 0.5 atoms cm$^{-3}$ is depicted in figure 2 where the grid surface shows the temporal variation of the CR energy density relative to the initial value for distances from 30-50 pc from the explosion center. After the shock wave has passed, the CR intensity decreases due to the adiabatic expansion of the remnant and the further evolution is then characterized by a diffusive transport of the accelerated particles from the shock wave towards the interior. For a remnant of radius R, the diffusion time scale can be estimated according to t ~ $R^2/K_{CR}$ yielding values between 270 and 750 kyr for 30 pc ≤ R ≤ 50 pc, respectively. In order to compute the CR intensity the SNR simulations have to be carried throughout the radiative cooling phase and it has been found that a SN at a distance of 40 pc increases the CR intensity only by about 15 %, however, for a period of some 100 kyr. In case of a distance of 30pc it will be already around 25%. Fig.2 shows the temporal variation of the cosmic ray energy density Ec relative to the initial value Ec,0 (100%) for distances from 30-50 pc from the explosion center. The calculations have been performed for an ISM density of 0.5 atoms cm. Less ISM density yields a lower CR intensity and vice versa.

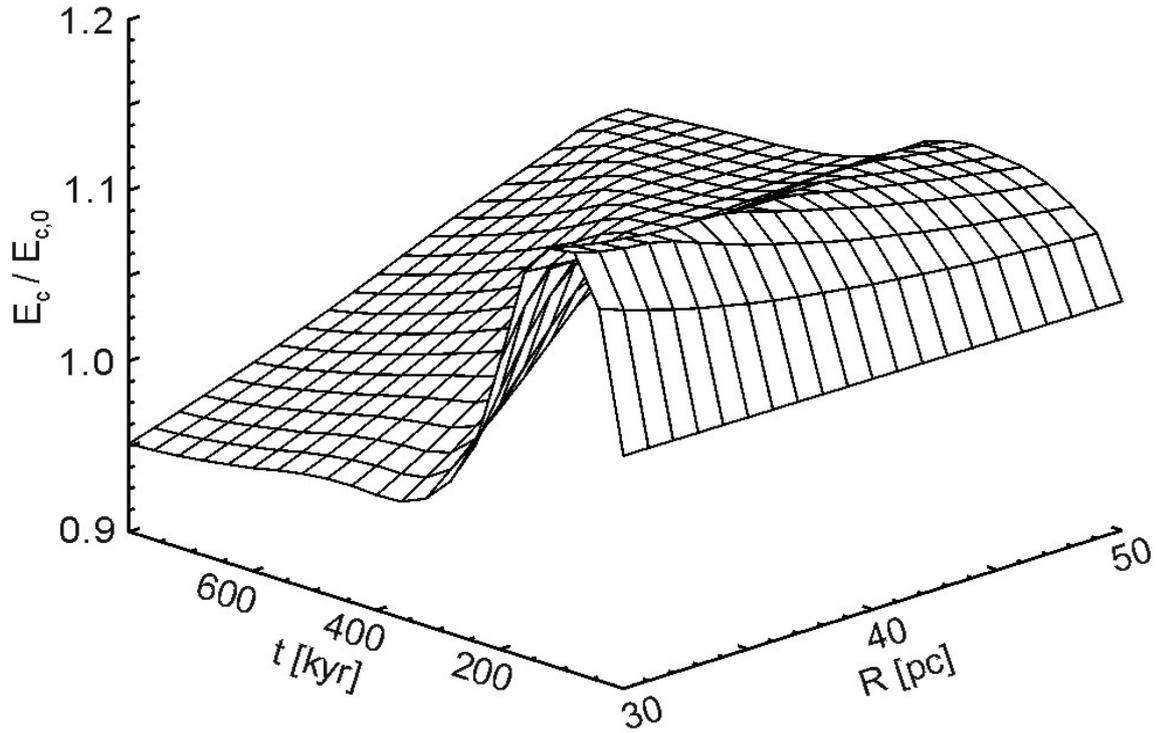

FIG. 2: Temporal variation of the cosmic ray energy intensity Ec relative to the initial value Ec,0 (100%) for distances from 30-50 pc from the explosion center. The calculations have been performed for an ISM density of 0.5 atoms cm$^{-3}$ "(Dorfi 1990)".

**2.2. Case B**

The situation is however different for a closer SN (case B).

In this case the pressure of the remnant $P_{ej}$ might overcame the solar wind ram pressure $P_{SW}$; this yields solar wind penetration.

$P_{ej} \approx (M_{tot} \times v)/(D^2 \times \Delta t) \geq P_{SW}$ ; $P_{SW} \approx m_H \times v_{SW} \times \phi_{SW}$; $v_{SW} \approx 400$ km/s; $\phi_{SW} \approx 3 \times 10^8$ H$^+$/(cm$^2 \times$ s)

$D_{crit} \approx 20$ pc

$M_{tot}$ total mass of remnant plus collected ISM ($\varsigma_{ISM} \approx 1$ cm$^{-3}$)

V ≈ 100km/s velocity of remnant; Δt ≈ duration; $\phi_{SW}$ solar wind flux

The transition from case A to case B is expected to occur roughly between 10pc and 30pc (transition from the Taylor-Sedov phase to the Snow-plow phase) but it depends also on the density ς of the interstellar medium.

Early studies on the impact to the Earth of such a close SN have been performed by "(Terry and Tucker 1968)", "(Rudermann 1975)" and rather elaborated studies have been performed later by J. Ellis et al. "(Ellis and Schramm 1995)", "(Ellis et al. 1996)". J.Ellis et al. considered a SN at a distance of ≈10pc. They estimated the fluence of neutral ionizing radiation at Earth's surface to be $\phi_n = 6.6 \times 10^5$ $(10/D_{pc})^2$ erg/cm$^2$ lasting for about a year. The charged CR fluence is estimated to be $\phi_{cr} = 7.4 \times 10^6 (10/D_{pc})^4$ erg/cm$^2$ over a duration of $\approx 3D^2_{pc}$ yr. $D_{pc}$ is the distance of the SN measured in pc. We note that the normal cosmic ray flux (in absence of SN events) at the top of the atmosphere is $9 \times 10^4$ erg/(cm$^2$yr) "(Terry and Tucker 1968)". Thus in case of a distance of $\approx$10pc for a SN, the cosmic ray flux is estimated to be two orders of magnitude above the normal cosmic ray flux.

## 3 Biological effects

### 3.1 Direct biological effects (case B)

Terry and Tucker "(Terry and Tucker 1968)" discussed the biological consequence for terrestrial life in case of an increased cosmic ray flux of more than one order of magnitude lasting, at most, a few days. Considering the arguments presented above, only a SN closer than $\approx$ 20pc could generate such a flux (scenario B). Because the observed mass extinctions of the fauna show, at the same time, little effect on the flora, the authors conclude that the explosion of a SN should be considered as one possible mechanism. From recent studies by J.Vives i Batlle "(J.Vives i Batlle 2012)", and taking 0.3 $\times 10^{-3}$Gy/year as normal cosmic ray intensity on Earth, the authors argue that an increase of about two orders of magnitude would be required for producing a harmful dosage: "For small mammals, dose rates $\leq 2 \times 10^{-2}$Gy/day are not fatal to the population; for larger mammals, chronic exposure at this level is predicted to be harmful". The effects are quite diverse because of the variations in radiation resistance of different specious. For example as female mice are sterilized at a certain dose, it needs an about 2 orders of magnitude higher dose to kill insects and single-cell organisms "(Terry and Tucker 1968)". In a similar way, plants are as well less affected than animals at high doses "(Terry and Tucker 1968)". Hence the argument, that little effect on the flora but mass extinctions at the fauna could be an indication for a threat by a SN.

### 3.2. Indirect effects on the biosphere

### 3.2.1 Ozone depletion (case B)

In a different approach from the previous discussion, Ellis and Schramm "(J. Ellis and D. Schramm 1995)" followed the discussion of Ruderman "(Rudermann 1975)" who studied the depletion of the ozone layer. Because of the enhancement of the cosmic ray intensity of more than one magnitude (scenario B) they estimate in the same way an increase of stratospheric production of NO. This is by a catalytic destruction of ozone due to two reactions "(Crutzen 1970)"

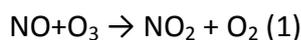
$NO + O_3 \rightarrow NO_2 + O_2$ (1)

and

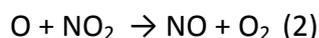
$O + NO_2 \rightarrow NO + O_2$ (2)

Ellis and Schramm concluded in their analyses an ozone hole depletion of roughly 95%, induced by a SN at a distance of 10pc, would last for $\approx$300yr. They focus on a disruption of the food chain, induced by the depletion of photosynthesizing organism. They describe consequences of a long-term exposure of the ultraviolet radiation: mass extinctions of marine life and also buildup of $CO_2$, yielding greenhouse episodes. However in a later publication "(Crutzen and Brühl 1996)" estimated a substantial smaller ozone depletion. Instead of 95% "(Ellis and Schramm 1995)", they estimated by using a detailed atmospheric model calculation a much smaller ozone depletion of at most 60% at high latitudes and below 20% at the equator. They conclude that a significant stress might have affected the biosphere at high latitudes only but much less on the tropical and subtropical biosphere, not strong enough to cause mass extinctions. Nevertheless we know that marine organisms and also the terrestrial ecosystems are affected by damages and also mutations by an increased exposure of UV radiation "(Smith 1972)". Marine organism in shallow water, suffer damage by increased UV radiation. Plants, animals and microorganisms are as well affected by increased UV radiation.

More recently, a very detailed study on ozone depletion has been performed by by Gehrels et al. "(Gehrels 2003)". They used a quite elaborated, two dimensional, photochemical transport model from NASA Goddard Space Flight Center. They concluded that a significant effect of ozone depletion due to the combined effect of gamma and cosmic rays from a SN, might occur by a SN closer than 8pc. However to get a clear picture is rather difficult. The reason is the enormous complexity of the biosphere and the uncertainties of the different parameters involved. Also Gehrels et al. "(Gehrels 2003)" point out, that only since between around 600Ma and 500Ma before present, the Earth has had developed enough oxygen in the atmosphere to form an ozone shield. Hence all the discussions above are only valid after this time. At any time before different scenarios between cosmic rays from a close SN and the atmosphere might have determined the climate at that time. Martin et al. "(Martin 2009)" discusses only short-term effects of gamma ray bursts on Earth covering mostly Archean and Proterozoic eons. It will be discussed below in the context of gamma ray bursts.

### 3.2.2. Cloud formation (case A and B)

It was Ney "(Ney 1959)" who first pointed out the existence of a large tropospheric and stratospheric effects produced by the solar-cycle modulation of cosmic rays confirmed by weather observations. He also suggested climatological effects due to cosmic rays. Later Svensmark "(Svensmark and Friis-Christensen 1995)", "(Svensmark 1998)" propagated also the idea that cloud formation follows the variations in GCR. It has also shown experimentally that ions play an important role for nucleation process in the atmosphere and, as they state, at the end for cloud formation "(Svensmark 2007)". The connection between cloud formation and the climate on Earth is well known. In their book, Svensmark and Calder "(Svensmark and Calder 2007)" postulate that ice ages might be caused by temporary increases in the intensity of CR. This is actually a proposition introduced by N.J. Shaviv "(Shaviv 2002)". Shaviv suggested that through the periodical crossing of the spiral arms in our Galaxy, the solar system is exposed also to a periodical increase of SNe and hence CR, since SNe tend to occur in the active star-forming regions in the spiral arms "(Dragicevich

1999)". Shaviv constructed a Galactic cosmic ray diffusion model and deduced a periodically increase in the CR flux associated with the travel of the solar system through the spiral arms of the Milky Way. In addition he compared his results with the exposure ages of iron meteorites and found correlations: Iron meteorites are usually exposed to CR for several 100My within of the solar system. By nuclear reactions on the elements in the meteoroid radionuclides and stable isotopes are formed, from which the exposure time can be deduced. A considerable correlation, postulated by Shaviv "(Shaviv 2002)", is seen between the cosmic ray fluctuation and ice-age epochs. Nevertheless there are still discussions on the unambiguously disentangling of Shaviv's arguments from the climate forcing induced by the, normally invoked, Croll-Milankovitch effect for glaciation cycles "(Beech 2011)", "(Williams 2003)". Also a detailed analyses of the exposure time of iron meteorites, considering also minor elements like S and P for the production of stable 21Ne (neglected by Shaviv), indicates that the variations mentioned by Shaviv might be less pronounced as thought before "(Ammon 2008)".

Although our understanding of the connection between cosmic rays and cloud formation is incomplete, this field of research is experimentally accessible at accelerator facilities able to simulate CR. The Cosmics Leaving Outdoor Droplets (CLOUD) (Cloud 2015) experiment at CERN studies the possible link between galactic cosmic rays and cloud formation "(Cloud 2015)". Based at the Proton Synchrotron (PS) at CERN, simulating cosmic rays, CLOUD aims to further understanding of aerosols and clouds, and unravelling possible connections between cosmic rays and clouds and their effect on climate. This might help to quantify any possible correlations between SN and climate and perhaps also mass extinctions.

**4. Frequency of close SNe**

Gehrels et al. "(Gehrels 2003)" estimated the frequency of close SN within the past several hundred million years. They considered a SN with total gamma-ray energy of $\approx 1.8 \times 10^{47}$ erg. For a critical distance, to disrupt Earth's ozone layer, of $D_{crit} \approx 8pc$ (case B), the time-averaged Galactic rate of close core-collapse SNe was calculated to be $\approx 1.5$ Ga$^{-1}$. A more frequent rate has been calculated by Beech "(Beech 2011)". Beech estimates, assuming random locations for SNe within the galactic disk, a time interval of the order of 112Ma. He considered SNe occurring within a critical distance of 10pc. He discusses also that most of the SNe occure mostly interior to the Sun's orbit about the galactic center which should change the time interval. Other effects, like the encounter with the spiral arm structure of the galaxy, are also discussed. His final conclusion is that the time interval should likely be between $\approx$75Ma and $\approx$150Ma. However we should keep in mind that the rate would go up significantly for more than one order of magnitude if distances like ~40pc would be included (case A). This average rate might, however, not represent a good estimate for near past/future, since this depends on the local composition of the solar neighborhood, such as possible encounters of the solar system with highly active star associations "(Breitschwerdt 2012).

## 5. Gamma ray bursts

A different scenario of a SN threat to life on Earth could be posed gamma ray bursts GRBs. In the electromagnetic spectrum, GRBs are the most energetic events in the universe, lasting only milliseconds to hours. This short timescale leads to the association of GRBs with explosion-like events such as SNe and neutron star mergers. At first gamma ray bursts have been observed by the US spacecraft Vela; initially installed in Earth's orbit for surveillance of nuclear bomb tests.

It is believed that the flash is released in two narrow cones, opposite to each other, a precondition which drastically reduce the probability of an interaction. Long lasting gamma ray bursts (>2s) are discussed in conjunction with core collapse of very massive stars which yield a neutron stare or a black hole. Hypothetical hypernovae yielding a black hole are as well considered.

Among others, one scenario being discussed is the astrophysical site of short lasting flashes (<2s) SN1 of type Ia. In this scenario, a white dwarf star accretes mass via Roche-Lobe overflow from a (typically) main sequence companion star until its mass exceeds the Chandrasekhar limit, causing gravitational collapse and subsequently, the SN explosion.

One of the first discussions with respect to terrestrial implications of gamma ray bursts have been performed by S.E. Thorsett "(Thorsett 1995)".

Quite detailed discussions have been performed by A.L. Melott et al. "(Melott 2003)", and especially by Thomas et al. "(Thomas 2005)", "(Thomas 2005)". They argued that gamma ray fluences of 10kJ/m², 100kJ/m², and 1000kJ/m² cause depletions of the ozone layer in a time range of one month of 68%, 91%, and 98% respectively. The reaction cycle they consider is the following: $NO_y$ compounds are created by the dissociation of $N_2$ due to the gamma rays in the stratosphere, by the reaction with $O_2$, NO is formed. Furthermore, $NO + O_3 \rightarrow NO_2$, $NO_2 + O_2$, $NO_2 + O \rightarrow NO + O_2$. The result is an increase in $NO_y$ and the decrease in $O_3$. From the ozone column densities they estimated the UVB flux at Earth's surface and the resulting DNA damage. The question what might have happened at Archean (~4000 Ma until ~2500 Ma) and Proterozoic (~2500 Ma until ~550 Ma) eons is discussed by O. Martin et al. "(Martin 2009)". They consider an atmospheric oxygen level between $10^{-5}$ of the present atmospheric level of $O_2$ to roughly 20% $O_2$ of modern Phanerozoic atmosphere. It should be considered that the $O_2$ rise in the Earth's atmosphere might be linked with the development of first photosynthetic bacteria "(Catling 2005)". The effect they study under these conditions is mostly from the reemission of interaction of primary γ- and X-ray radiation with the atmosphere at that time. Because of the smaller oxygen content a greater DNA damaged is expected. Critical distances for lethality are calculated to be in the order of ~200 pc. Because of lack of clear indicators from these eons this discussion is still speculative.

How frequent are GRBs which might have hit the Earth in the past? Beech "(Beech 2011)" estimated the frequency of GRBs being closer than 1kpc to the Sun. He assumed that about

1 in 5000 SNe form a hypernova which could cause a GRB, radiating in two highly focused, narrow opposing cones. His calculations show a very low rate of roughly 1 in 2Ga causing threat on Earth.

**6. Candidates of nearby SNe**

What are at possible candidates for a nearby SN at present? Beech addressed this question in a detailed study "(Beech 2011)". There a two possible candidates, IK Pegasi and Betelgeuse. IK Pegasi is a binary star system, an A8 spectral type main sequence star and a DA white dwarf. It is considered as the closed system to the Sun which will generate in the end a Ia SN "(Parthasarathy 2007)". Beech estimates that the system will be at the closest distance of (40±2) pc in (1.1±0.1) Myr. According to his estimations IK Pegasi will become a Type Ia SN around 1.9Gy from now. The system than will be many kiloparsec away from the Sun, posing no threat to the Earth's biosphere.

The other candidate is Betelgeuse, a M2Iab supergiant. It is located around 200pc away from the Sun and moving away from our solar system with a relative velocity of about 33km/s. According to Beech "(Beech 2011)" it is expected that Betelgeuse undergo core-collapse at any time within the next 2 million years. However because of the distance of around 200pc and the discussions above we shall not expect any threat from Betelgeuse when it explodes once as a Sn Type II.

**Conclusions**

Mass extinction events can have very different causes like palaeoclimatic, palaeoecologic, and also palaeoenvironmental explanations. There are often inconsistences which contradict one clear explanation only. Beside of the big five extinctions there are quite a few smaller extinctions. Even that clear pattern do not exist, however because of the frequency of nearby SNe or even GBRs, which are much less frequent, one or more of these events might have initiated mass extinction(s) in the past.

**Acknowledgement**

The critical comments of Dr. P. Ludwig on this manuscript are greatly acknowledged.